\documentclass[reprint,aps,pra, superscriptaddress, longbibliography]{revtex4-2}
\usepackage{amsmath}
\usepackage{amssymb}
\usepackage{graphicx}

\usepackage[
    colorlinks,
    linkcolor=blue,
    anchorcolor=black,
    citecolor=blue,
    urlcolor=blue,
]{hyperref}



\begin{document}
\title{Cold hybrid electrical-optical ion trap}
\author{Jin-Ming Cui}
\email{jmcui@ustc.edu.cn}
\affiliation{Laboratory of Quantum Information, University of Science and Technology of China, Hefei 230026, China}
\affiliation{Anhui Province Key Laboratory of Quantum Network, University of Science and Technology of China, Hefei 230026, China}
\affiliation{CAS Center for Excellence in Quantum Information and Quantum Physics,
University of Science and Technology of China, Hefei 230026, China}
\affiliation{Hefei National Laboratory, University of Science and Technology of
China, Hefei 230088, China}

\author{Shi-Jia Sun}
\affiliation{CAS Key Laboratory of Quantum Information, University of Science and Technology of China, Hefei 230026, China}
\affiliation{CAS Center for Excellence in Quantum Information and Quantum Physics,
University of Science and Technology of China, Hefei 230026, China}

\author{Xi-Wang Luo}
\email{luoxw@ustc.edu.cn}
\affiliation{CAS Key Laboratory of Quantum Information, University of Science and Technology of China, Hefei 230026, China}
\affiliation{Anhui Province Key Laboratory of Quantum Network, University of Science and Technology of China, Hefei 230026, China}
\affiliation{CAS Center for Excellence in Quantum Information and Quantum Physics,
University of Science and Technology of China, Hefei 230026, China}
\affiliation{CAS Center for Excellence in Quantum Information and Quantum Physics,
University of Science and Technology of China, Hefei 230026, China}

\author{Yun-Feng Huang}
\affiliation{CAS Key Laboratory of Quantum Information, University of Science and Technology of China, Hefei 230026, China}
\affiliation{Anhui Province Key Laboratory of Quantum Network, University of Science and Technology of China, Hefei 230026, China}
\affiliation{CAS Center for Excellence in Quantum Information and Quantum Physics,
University of Science and Technology of China, Hefei 230026, China}
\affiliation{Hefei National Laboratory, University of Science and Technology of
China, Hefei 230088, China}

\author{Chuan-Feng Li}
\email{cfli@ustc.edu.cn}
\affiliation{CAS Key Laboratory of Quantum Information, University of Science and Technology of China, Hefei 230026, China}
\affiliation{Anhui Province Key Laboratory of Quantum Network, University of Science and Technology of China, Hefei 230026, China}
\affiliation{CAS Center for Excellence in Quantum Information and Quantum Physics,
University of Science and Technology of China, Hefei 230026, China}
\affiliation{Hefei National Laboratory, University of Science and Technology of China, Hefei 230088, China}

\author{Guang-Can Guo}
\affiliation{CAS Key Laboratory of Quantum Information, University of Science and Technology of China, Hefei 230026, China}
\affiliation{Anhui Province Key Laboratory of Quantum Network, University of Science and Technology of China, Hefei 230026, China}
\affiliation{CAS Center for Excellence in Quantum Information and Quantum Physics,
University of Science and Technology of China, Hefei 230026, China}

\date{\today}
\begin{abstract}
    Advances in research such as quantum information and quantum chemistry
    require subtle methods for trapping particles (including ions, neutral
    atoms, molecules, etc.). Here we propose a hybrid ion trapping method
    by combining a Paul trap with optical tweezers. The trap combines
    the advances of the deep-potential feature for the Paul trap and the
    micromotion-free feature for the optical dipole trap. By modulating
    the optical-dipole trap synchronously with the radio frequency voltage of the Paul trap,
    the alternating electrical force in the trap center is fully counteracted, and
    the micromotion temperature of a cold trapped ion can reach the order
    of nK while the trap depth is beyond 300K. These features will enable
    cold collisions  between an ion and an atom in the
    $s$-wave regime and stably trap the produced molecular ion in the
    cold hybrid system. 
    This will provide a unique platform for probing the interactions between the
    ions and the surrounding neutral particles and enable the investigation
    of new reaction pathways and reaction products in the cold regime.
\end{abstract}
\maketitle
\section{Introduction} 
Near absolute zero temperatures, cold chemistry unveils intricate quantum processes within reaction mechanisms, garnering significant attention across both chemical and physical disciplines \cite{heazlewood2021towards, he2020coherently, hu2019directobservation, KKN2020PRL, KKN2020PRR, KKN2022AnnualReview, Bo2022nature, Bo2022prl, Bo2023pra, Cheuk2020PRL, Xie2020Science, Jongh2020Science}.
In the last decade, an emerging area of cold hybrid ion-atom systems
has been developed as a new platform for fundamental research, 
which is dedicated to controlling the cold collision and quantum chemistry
of ions and atoms \cite{harter2014coldatomtextendashion, tomza2019coldhybrid, 
zuber2022observation, zou2023observation, weckesser2021observation, 
hirzler2022observation, katz2022quantum}.
Binding the neutral atom and ion and then stably trapping the produced molecular
ion are prerequisites for exploring the quantum features of the molecular ion. 
To investigate collision physics in the quantum regime 
of the cold hybrid ion-atom system, reaching the single
partial wave limit (i.e., the $s$-wave limit) is an important step \cite{tomza2019coldhybrid}.
Although the trapped ion can be cooled close to the quantum collision regime using buffer atomic 
gases with a large ion-atom mass ratio \cite{feldker2020buffergas},
the micromotion of the Paul trap (PT) heats the system away from a
stable bound state \cite{cetina2012PRL,pinkas2023NP,hirzler2023trapassisted},
which hinders progress in this area. Pioneering experiments using optical dipole traps
(ODTs) to capture ions were demonstrated \cite{schneider2010optical,huber2014afaroffresonance,schmidt2018optical,schmidt2020optical},
which is a potential solution to the micromotion heating problem in
the cold hybrid ion-atom system. However, because the dipole potential is shallow, 
it is quite challenging to keep the ion during the reaction and 
stably trap the produced molecular ion.

In this work, we propose a novel hybrid electrical-optical ion trap that combines 
the advantages of the deep-potential feature of a PT
and the micromotion-free feature of an ODT. 
The key idea is to counteract the alternating electrical force around the center of a PT 
that is responsible for the micromotion problem, 
which is achieved by introducing optical-dipole
traps that are modulated synchronously with the radio frequency (RF) voltage of the PT.
This hybrid electrical-optical trap corresponds to a deep PT with a micromotion-free dip at the center, 
and this potential structure provides an ideal platform for studying low-energy physics 
(such as cold collisions and cold chemistry). 
Since  the  center dip provides a micromotion-free area for the trapped ion to interact with a neutral atom, 
the produced molecular ion can be stably kept in a deep background potential profile provide by the PT. 
To analyze the performance of the trap, we consider
counteracting $99\%$ of the alternating force which is experimentally reasonable. 
The intrinsic micromotion (iMM) energy can be suppressed by four orders of magnitude. 
Under current voltage compensation techniques
in trapped ion fields \cite{gloger2015iontrajectory}, 
the DC stray field can be suppressed below 1~V/m, 
which leads to the ultimate cold ion excess micromotion (eMM) temperature
in the hybrid trap below 1~nK. 
Under such cold ion temperatures, many ion-atom $s$-wave scattering process can be investigated.
Moreover, by simulating a cold collision process between a $\mathrm{Yb^{+}}$
ion and a Rb atom, we find that
the hybrid trap effectively solves the micromotion heating problem,
which dissociates the ion-atom bound state in the cold collision process. 
The proposed hybrid trap offers a simple yet powerful platform to study ion-atom collisions in the quantum regime,
which paves the way for exploring novel reactions and manipulating the produced  molecular ion in such cold hybrid systems.

\section{Scheme and Results}
Our scheme employs modulating optical tweezers to counteract the alternating force
caused by the RF electrical field. 
If we consider a hybrid trap consisting of a PT and $N$ modulating ODTs, 
the total potential of the trap is 
\begin{align}
    U(\mathbf{r},t)= & U_{E}(\mathbf{r})+\widetilde{U}_{E}(\mathbf{r})\cos(\Omega t)\nonumber \\
    & +\sum_{i}^{N}U_{D,i}(\mathbf{r})\left[1+\eta_{i}\cos(\Omega t)\right],\label{eq:U_total}
\end{align}
where $U_{E}(\mathbf{r})$ and $\widetilde{U}_{E}(\mathbf{r})$ are
the potentials generated by the DC field and RF field of the PT respectively, 
$U_{D,i}(\mathbf{r})$ is the potential of $i$'th
ODT, and $\eta_{i}$ is the modulation depth of $i$'th ODT with $\left|\eta_{i}\right|\leq1$. 
When the ion is cooled to a low temperature (e.g., via Doppler cooling to several millikelvin), its motional range becomes significantly smaller than the characteristic size of the Paul trap and the optical dipole trap (ODT).
Supposing the centers of all traps overlap each other for simplicity, 
the potentials around the trap center can be expanded as harmonic forms
\begin{align}
    U_{E}(\mathbf{r}) & =k_{0}^{x}x^{2}+k_{0}^{y}y^{2}+k_{0}^{z}z^{2} + C_E,\\
    \widetilde{U}_{E}(\mathbf{r}) & =k_{1}^{x}x^{2}+k_{1}^{y}y^{2}+k_{1}^{z}z^{2} + \widetilde{C}_{E},\\
    U_{D,i}(\mathbf{r}) & =K_{i}^{x}x^{2}+K_{i}^{y}y^{2}+K_{i}^{z}z^{2} + C_{D,i},
\end{align}
where $C_E$, $\widetilde{C}_{E}$ and $C_{D,i}$ are constants.  $U_{E}(\mathbf{r})$ and $\widetilde{U}_{E}(\mathbf{r})$ generated
by electrical fields have to fulfill the Laplace equation $\Delta U(\mathbf{r})=0$,
which leads to restrictions in the geometric factors,
\begin{align}
    k_{0}^{x}+k_{0}^{y}+k_{0}^{z} & =0,\label{eq:DC_restric}\\
    k_{1}^{x}+k_{1}^{y}+k_{1}^{z} & =0,\label{eq:RF_restrict}
\end{align}
while $U_{D,i}(\mathbf{r})$ is not restricted by this condition.
The alternating force strength of the hybrid trap is
$\mathbf{\widetilde{F}(\mathbf{r})}=\nabla\left(\widetilde{U}_{E}+\sum \eta_i U_{D,i}\right)$, so eliminating the force to zero leads to the condition
\begin{equation}
    k_{1}^{\gamma}+\sum_{i}^{N}\eta_{i}K_{i}^{\gamma}=0,\label{eq:condition}
\end{equation}
for all $\gamma\in\left\{ x,y,z\right\} $, which can be achieved by setting the modulation
depths and geometric shapes of ODTs. 
\begin{figure}
    \centering\includegraphics[width=1\columnwidth]{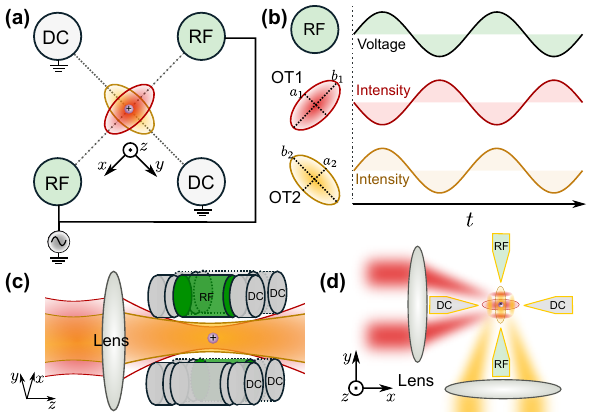}
    \caption{\label{fig:Scheme}
    Scheme of a hybrid electrical-optical ion trap.
    (a) The trap consists of a Paul trap (e.g., a four-rod trap or a
    blade trap) and two modulating optical dipole traps (ODTs) with different
    oriented astigmatic spots. (b) The phase relation between the RF
    voltage of the Paul trap and the optical intensities of the two ODTs.
    (c,~d) two feasible realizations of the hybrid trap.}
\end{figure}

Supposing we use astigmatic Gaussian spots to generate the dipole
traps required. The intensity profile of a focused astigmatic Gaussian
beam along axis $c$ in coordinate $(\hat{a},\hat{b},\hat{c})$ can
be written as
\begin{equation}
    I(\mathbf{r})=I_{0}\frac{w_{0a}}{w_{a}(c)}\frac{w_{0b}}{w_{b}(c)}\exp\left\{ -2\left[\frac{a^{2}}{w_{a}(c)^{2}}+\frac{b^{2}}{w_{b}(c)^{2}}\right]\right\} ,\label{eq:I}
\end{equation}
where $w_{a/b}(c)=w_{0,a/b}\sqrt{1+(c/z_{Ra/b})^{2}}$ are the beam
waists for $a$ and $b$ axes, i.e., short and long axes respectively,
$z_{Ra/b}=\pi w_{0,a/b}^{2}/\lambda$ are the Rayleigh ranges and
$\lambda$ is the wavelength. As the potential of ODT is proportional
its intensity $U_{D,i}(\mathbf{r})\propto I_{D,i}(\mathbf{r})$, by
expanding it near the zero point, the potential coefficients are
\begin{align}
    K_{i}^{a} & =\frac{2U_{0i}}{w_{0,ai}^{2}},K_{i}^{b}=\frac{2U_{0i}}{w_{0,bi}^{2}},\nonumber \\
    K_{i}^{c} & =\frac{U_{0i}\lambda^{2}}{2\pi^{2}}\left(\frac{1}{w_{0,ai}^{4}}+\frac{1}{w_{0,bi}^{4}}\right),\label{eq:Kabc}
\end{align}
where $U_{0i}=\frac{\pi c^{2}\Gamma}{2\omega_{0}^{3}}(\frac{2}{\Delta_{2}}+\frac{1}{\Delta_{1}})I_{0i}$
\cite{grimm2000optical}. Finding a proper set of $\left\{ \eta_{i}U_{0i},\,w_{0,ai},\,w_{0,bi}\right\} $
to satisfy Eq.~\ref{eq:condition} and Eq.~\ref{eq:Kabc} will eliminate
the alternation force at the trap center .

Figure~\ref{fig:Scheme}(c,d) shows two feasible configurations
to demonstrate the scheme in a symmetric linear PT with four
rods or blades \cite{berkeland1998minimization,drewsen2000harmonic}.
For simplicity, we take Fig.~\ref{fig:Scheme}(c) for example, where
two optical tweezers with focused astigmatic Gaussian spots are applied
on the trapped ion with similar beam waists, i. e., $w_{0,a1}=w_{0,a2}=w_{a}$,
$w_{0,b1}=w_{0,b2}=w_{b}$, $U_{01}=U_{02}=U_{0}$, and the 
long axes of two two ODTs are perpendicular to each other, i.e. $\hat{a}_{1}\perp\hat{a}_{2}\parallel\hat{y}$,
$\hat{b}_{1}\perp\hat{b}_{2}\parallel\hat{x}$. The symmetric linear
PT condition requires $k_{1}^{x}=-k_{1}^{y}=k_{1}$, $k_{1}^{z}=0$.
Suppose $\eta_{1}=-\eta_{2}$=1, which indicates modulations between
two ODTs have a $\pi$ phase difference (shown in Fig.~\ref{fig:Scheme}(b)),
then the configurations of the ODTs can be 
\begin{align}
    k_{1} & =k_{a}-k_{b},\nonumber \\
    K_{1}^{x} & =K_{2}^{y}=k_{a},\nonumber \\
    K_{1}^{y} & =K_{2}^{x}=k_{b},\label{eq:K_relation}
\end{align}
where $k_{a,b}=2U_{0}/w_{a,b}^{2}$.

\begin{figure}
    \centering\includegraphics[width=0.95\linewidth]{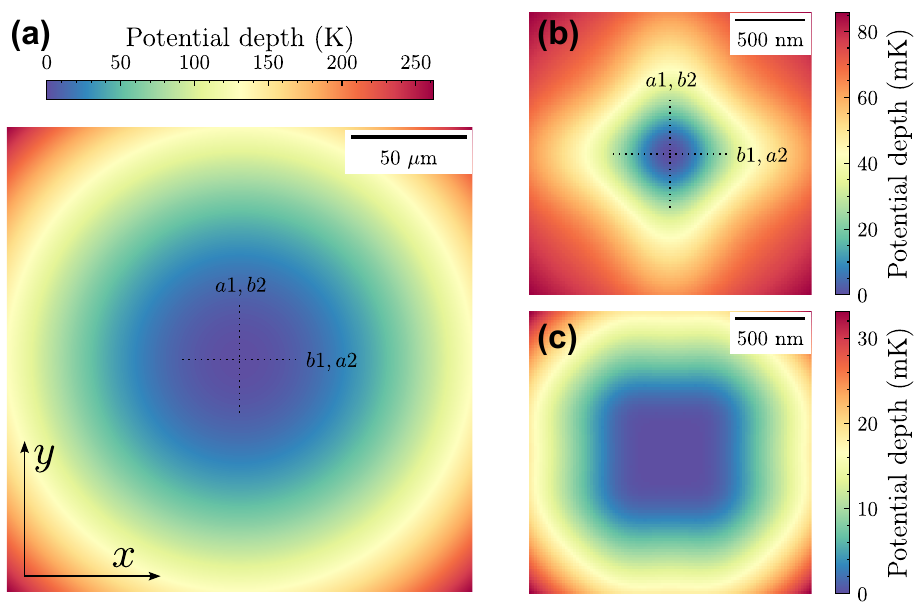}
    \caption{\label{fig:PseudoP}
    Pseudo potential of a hybrid trap for $\mathrm{^{171}Yb^{+}}$.
    (a) The total pseudo potential of the trap, indicating a deep potential.
    (b,c) The constant part and the alternating part of the pseudo potential
    near the trap center, respectively. The alternating pseudo potential
    is flat near the center, which indicates the alternating force is
    zero. The simulation condition is $\Omega=2\pi\times2$\ MHz, $V_{RF}=10$\ V,
    $\kappa=0.6$, $V_{DC}=1$\ V, $R=0.5$\ mm, $Z_{0}=1$\ mm, $\omega_{x}=\omega_{y}=2\pi\times180$\ kHz,
    $\omega_{z}=2\pi\times131$\ kHz for the Paul trap, laser wavelength
    at 554.6\ nm, $w_{a}=w_{b}/3=478$\ nm can be generated through
    an objective with a numerical aperture (NA) of 0.6, $U_{0}/h=-622$\ MHz
    for optical dipole traps.}
\end{figure}
The pseudo potential of the hybrid trap can be decomposed into a direct
potential (DP) and an alternating potential (AP)
\begin{equation}
    \varPsi(\mathbf{r})=\Phi_{0}(\mathbf{r})+\frac{1}{4m\Omega^{2}}\left\Vert \nabla\Phi_{1}(\mathbf{r})\right\Vert ^{2},\label{eq:Pseudo}
\end{equation}
where $\Phi_{0}(\mathbf{r})=U_{E}(\mathbf{r})+\sum_{i}^{N}U_{D,i}(\mathbf{r})$
and $\Phi_{1}(\mathbf{r})=\widetilde{U}_{E}(\mathbf{r})+\sum_{i}^{N}\eta_{i}U_{D,i}(\mathbf{r})$
are the direct and alternating potentials respectively. For a symmetric
linear PT \cite{berkeland1998minimization}, $k_{0}^{x}=k_{0}^{y}=-2k_{0}^{z}=-2\kappa QV_{DC}/Z_{0}^{2}$,
$k_{1}^{x}=-k_{1}^{y}=QV_{RF}/2R^{2}$, $k_{1}^{z}=0$, where $\kappa$
is a geometric factor of the trap, $Q$ is the charge of the ion ,
$V_{DC}$ and $V_{RF}$ are the DC and RF voltages applied to the trap,
$R$ is the perpendicular distance from the trap axis to the trap
electrodes, $Z_{0}$ is the distance from the trap center to the endcap
electrodes. The pseudo potential can be calculated by substituting
Eq.~\ref{eq:I}-\ref{eq:K_relation} into Eq.~\ref{eq:Pseudo}.
When the alternating force caused by the RF electrical field is completely
compensated by the modulating ODTs, which we call the zero alternating
trapping condition (ZATC), we can derive $U_{0}=QV_{RF}w_{a}^{2}w_{b}^{2}/4R^{2}(w_{a}^{2}-w_{b}^{2})$.
To make the condition realistic, a low RF voltage of PT
and small spots of ODTs should be used to minimize the optical power
of the ODT. Consequently, a lower RF frequency $\Omega$ is preferred
to make a deeper pseudo potential under the low RF voltage. A numerical
simulation result is show in Fig.~\ref{fig:PseudoP}. In a large
range, it acts as a PT with deep potential as Fig.~\ref{fig:PseudoP}(a).
At the center of the trap, it acts as a tight ODT as Fig.~\ref{fig:PseudoP}(b).
The alternating pseudo potential in Fig.~\ref{fig:PseudoP}(c)
is flat, which indicates the alternating force is zero. With the
simulation result, the required laser power is 0.7 W at 554.6 nm,
if we use an objective with NA=0.6 to generate the required ODTs.

\begin{figure}
    \includegraphics[width=0.95\linewidth]{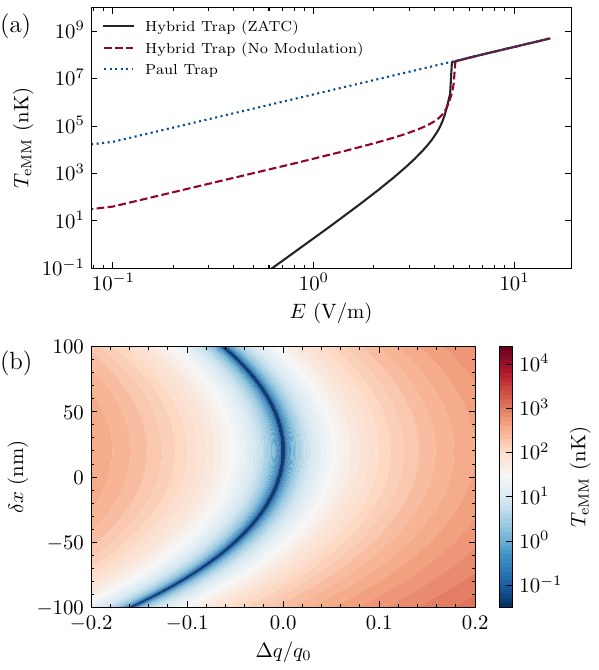}
    \caption{\label{fig:Temperature}
    Ion motion simulation for trapping temperature in a stray electrical field (SEF).  
    (a) Excess micromotion (eMM) temperature of a trapped
    ion along with different SEFs.
    The hybrid trap provides a lower temperature than a PT under
    the same SEF. When the hybrid trap works at zero
    alternating trapping condition (ZATC), the ion's eMM temperature 
    reaches nK under a SEF of 1 V/m. 
    The direction of the SEF for simulation is (1,1,1).
    (b) Ion's eMM temperature simulation under
    experimental imperfections, $\delta x$: misalignment between Paul trap and the
    ODT, $\Delta q$: uncompensated alternating force factor. Intrinsic mircomotion (iMM) is also supressed, as the iMM temperature is scale as ${\Delta q}^2$.
    The figure is calculated under a SEF of 1 V/m  along the $x$ axis.
    Hybrid trap and Paul trap parameters are the same as those in Fig.~\ref{fig:PseudoP}.
    }
\end{figure}
Next, we will discuss the cold trapped ion temperature limit in the
hybrid trap. Supposing the ion oscillates near $\mathbf{u}_{0}$ in
the hybrid trap under a stray electrical field  $\mathbf{E}$, and
has been cooled to a low temperature $\left|\mathbf{u}-\mathbf{u}_{0}\right|\ll w_{a}$,
then the motion equation can be approximated as a Mathieu equation
near $\mathbf{u}_{0}$
\begin{equation}
    \ddot{u_{l}}+\left[a_{l}(\mathbf{u}_{0})+2q_{l}(\mathbf{u}_{0})\cos\left(\Omega t\right)\right]\frac{\Omega^{2}}{4}u_{l}=\frac{Q\cdot E_{l}}{m},\label{eq:DCmotion}
\end{equation}
where $a_{l}(\mathbf{u})=4\hat{e_{l}}\cdot\nabla\Phi_{0}(\mathbf{u})/m\Omega^{2}u_{l}$
and $q_{l}(\mathbf{u})=2\hat{e_{l}}\cdot\nabla\Phi_{1}(\mathbf{u})/m\Omega^{2}u_{l}$.
The equilibrium position $\mathbf{u}_{0}$ can be numerically solved
by iterating
\begin{equation}
    u_{0,l}=\frac{4Q\cdot E_{l}}{m\Omega^{2}\left[a_{l}(\mathbf{u})+q_{l}^{2}(\mathbf{u})/2\right]}.\label{eq:u0}
\end{equation}
The final temperature caused by excess micromotion is 
\begin{equation}
    T_{\mathrm{eMM}}=\frac{m\Omega^{2}}{16k_{B}}\sum_{l}q_{l}^{2}(\mathbf{u}_{0})u_{0,l}^{2}.\label{eq:T_M}
\end{equation}
Using the above numerical method, we can study the ion eMM temperature
under a large electrical field. Fig.~\ref{fig:Temperature}(a) shows
the comparison of trapped ion eMM temperature in three different traps
under a stray electrical field along the (1,1,1) direction. The stray
field will shift the trap center of the PT part of the hybrid trap
while the ODT is not affected, so the total pseudo potential is a
double well under a small stray field. The ion stays in the tight
ODT under a small stray field, and the ion's eMM temperature will be low
even without modulating the ODTs. By increasing the stray field, 
the ion will eventually be pushed from the ODT to PT, 
so the ion's eMM temperature of the hybrid trap is close to that of the PT, 
when the stray field is larger than 5~V/m. 

\begin{table}
    \begin{ruledtabular}
        \caption{\label{tab:Wavelengths} ODT laser wavelengths for hybrid ion traps in different ion-atom systems.  
        \( E_k \) is a characteristic energy scale of the ion-atom interaction \cite{tomza2019coldhybrid}, which is associated with the \( s \)-wave scattering limit. The wavelengths are chosen to ensure that the alternating ODT at  a magic wavelength for the atom, minimizing the AC Stark shift of the atomic ground state. This prevents the ODT from driving the atom into micromotion as it approaches the ion. }
        \begin{tabular}{ccccc}
            Ion & Atom & $\lambda_{\mathrm{ODT}}$ (nm) & $P$ (W) & $E_{k}$/$k_{B}$ (nK)\tabularnewline
            \hline 
            $\mathrm{Yb^{+}}$ & Rb & 420.6  & 0.24 & 44.7  \\
            $\mathrm{Yb^{+}}$ & Rb & 422.3  & 0.25 & 44.7  \\
            $\mathrm{Yb^{+}}$ & Rb & 787.4  & 1.48 & 44.7  \\
            $\mathrm{Yb^{+}}$ & Yb & 554.63 & 0.70 & 44.7  \\
            $\mathrm{Ca^{+}}$ & Na & 589.46 & 0.48 & 1370.7\\
            $\mathrm{Ca^{+}}$ & Li & 670.97 & 1.09 & 10601.5\\
            $\mathrm{Sr^{+}}$ & Rb & 787.41 & 0.94 & 77.8  \\
        \end{tabular}
    \end{ruledtabular}
\end{table}

In order to further insight the ion eMM temperature relation with
experiment parameters (e.~g., misalignment between PT and
the ODT $\delta x$, stray electrical field $E$, residual alternating
force (RAF) parameter $\Delta q=4(k_{1}+k_{a}-k_{b})/m\Omega^{2}$
), we study one dimensional motion for simplification. Considering
the condition of small misaligned distance $\delta x\ll w_{0}$, the
approximated ion eMM temperature can be derived as 
(see Appendix B)
\begin{equation}
    T_{\mathrm{eMM}}\approx\frac{m\Omega^{2}}{16k_{B}}\left(\Delta q+\frac{(\delta x-x_{0}){}^{2}}{x_{b}^{2}}\right)^{2}x_{0}^{2},
\end{equation}
where $x_{0}\thickapprox-Q\cdot E/(k_{0}+2k_{a}+2k_{b})$ is the ion
position offset pushed by the stray electrical field, and $x_{b}^{2}=m\Omega^{2}/8(k_{b}/w_{b}^{2}-k_{a}/w_{a}^{2})$.
Fig.~\ref{fig:Temperature}(b) demonstrates that the ion eMM temperature can reach the 100~nK level, even with a misalignment of approximately 50~nm and a RAF ratio of $\Delta q/q_0 = 5\%$ (where $q_0 = 4k_{1}/m\Omega^{2}$ is the $q$-factor of the bare Paul trap), under a stray field of 1~V/m along the $x$-axis. 
Single-particle localization microscopy has demonstrated sub-10 nm positioning accuracy in ion imaging systems \cite{qian2021superresolved,blums2018asingleatom,wong-campos2016highresolution}. By implementing a dual-objective configuration on a high-optical-access ion trap \cite{he2021ion}—where opposing objectives are utilized with one generating optical tweezers and the other simultaneously imaging both ions and tweezer spots—we can achieve co-localization accuracy of less than 50 nm. This approach not only significantly reduces alignment complexity but also relaxes the trap alignment requirements in our scheme.
Additionally, the ion iMM temperature and energy, which scale as ${\Delta q}^2$ (see Appendix B), can be reduced by four orders of magnitude with a $1\%$ RAF ratio. The iMM temperature for the motional ground state of the bare Paul trap, calculated using the parameters in Fig.~2, is 2.73~$\mu$K. In contrast, for the hybrid trap with a RAF ratio of \(1\%\), it is significantly reduced to 58.3~pK.

\begin{figure}
    \includegraphics[width=1\columnwidth]{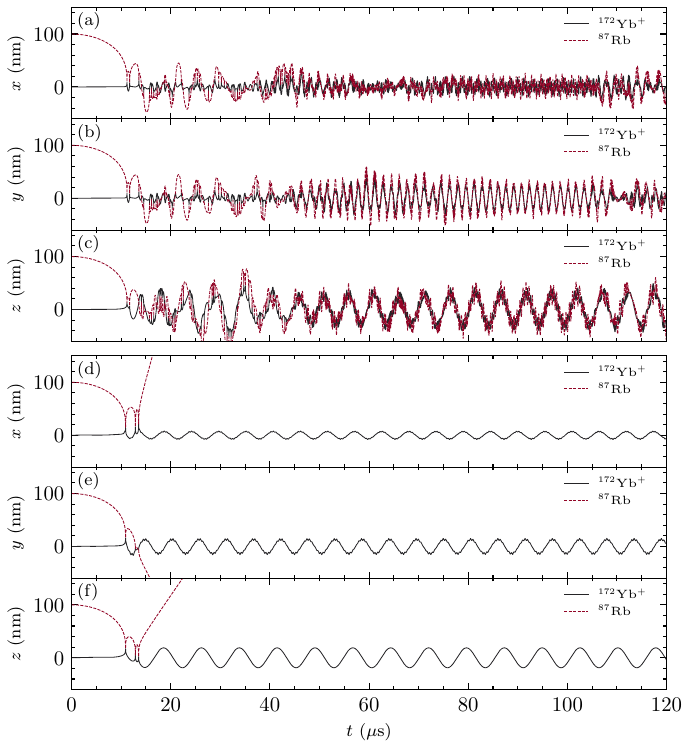}
    \caption{\label{fig:collison}
    Numerical simulations of a ion-atom (Rb and $\mathrm{Yb^{+}}$)
    collision in a hybrid trap (a-c) and a bare Paul trap (d-f). For comparison,
    the hybrid trap has the same electric fields with configurations as
    Fig.~\ref{fig:PseudoP}. Rb approaching an $\mathrm{Yb^{+}}$ initially
    stationed in trap center, the first hard-core collision time is at
    $t=0$. In Paul trap, the ion is heated by the RF field and bounces
    out quickly, whereas in hybrid trap, the ion-atom can be weakly bounded.
    The harmonic frequencies of two traps are $\omega_{\mathrm{Hybrid}}=2\pi\times0.84$~MHz
    and $\omega_{\mathrm{Paul}}=2\pi\times0.18$~MHz, leading to the
    corresponding characteristic lengths $R_{\mathrm{Hybrid}}=33.4$~nm
    and $R_{\mathrm{Paul}}=55.9$~nm, respectively. The simulation method
    is modified from Ref.~\citealp{cetina2012PRL}. }
\end{figure}
Finally, we will investigate the collision process between an ion and an atom in the hybrid trap. For this study, the laser wavelength of the ODT must be carefully selected based on the specific ion-atom pair. The alternating ODT for the trapped ion should not contribute to the trapping potential for the atom. This requires the ODT wavelength to be chosen such that it does not create a trapping potential for the atom, ensuring that the AC Stark shift of the atomic ground state is minimized. If this condition is not met, the alternating ODT would drive the atom into micromotion as it approaches the ion, resulting in unwanted dynamics. The wavelengths listed in Tab.~\ref{tab:Wavelengths} are carefully chosen to satisfy this requirement while balancing the tradeoff between ODT power and detuning from atomic and ionic resonances.
Tab.~\ref{tab:Wavelengths}
lists several laser wavelengths for some composition of ion-atom species,
and the laser power with hybrid trap setting as Fig.~\ref{fig:PseudoP}.  
We take Rb and $\mathrm{Yb^{+}}$ in this investigation following
the method presented in Ref.~\cite{cetina2012PRL},
with details in Appendix C.  When an ion moves in a PT without any disturbance, 
the RF field does zero work on the ion for an RF period. 
However, in the cold hybrid ion-atom system, the polarized
atom will exert an attraction force on the ion (by attraction potential
$V(r)=-C_{4}/r^{4}$ ) and thus disturb the conserved cycle of energy
transfer, and finally the ion-atom system gains net energy from the
RF field, which increases the temperature of the system. A characteristic
length scale $R=\sqrt[6]{2C_{4}/m_{i}\omega^{2}}$ is preferred to
be used (when the interaction potential $V(r)$ is equal in magnitude
to the trap harmonic potential, where $\omega=\Omega\sqrt{a+q^{2}/2}/2$
is the secular frequency of the ion near the trap center). In a PT
(Fig. \ref{fig:collison}(b)), the ion gains enough energy from the
RF field when the atom approaches, and after one collision, the atom
escapes. In the hybrid trap (Fig. \ref{fig:collison}(a)), since
the alternating force induced by the RF field near the trap center is compensated, 
the ion-atom system
nearly undergoes an energy-conservative process and forms a weakly
bound ion-atom pair, keeping a low temperature.
In the hybrid trap,
the corresponding characteristic energy scale is no longer restricted
by the RF heating but determined by the atomic gas temperature, which
enables access to the regimes of the $s$-wave limit or quantized ion
motion in the cold hybrid system. 
The work done by the RAF during
a collision depends on the phase $\phi$ at the time of the close-range
collision, which can be written as (see Appendix D)
\[
    W=\frac{1}{2}\mu\Omega^{2}R^{2}\sin\phi\cdot\left(\Delta q\eta_{1}+\frac{R^{4}}{r_{0}^{2}x_{b}^{2}}\eta_{2}\right)
\]
where $\mu=m_{i}m_{a}/(m_{i}+m_{a})$ is the reduced mass, $\eta_{1}=2/3\intop_{0}^{\Omega t_{1}}(A-B\tau^{1/3})\cdot\tau^{-2/3}\sin\tau\cdot d\tau$
and $\eta_{2}=2/3\int_{0}^{\Omega t_{1}}(A-B\tau^{1/3})^{3}\cdot\tau^{-2/3}\sin\tau\cdot d\tau$
are integrates with parameters $A=r_{c}r_{0}/R^{2}$, $B=\mu r_{0}^{2}/m_{i}R^{2}$,
the ion\textquoteright s displacement from the trap center
 $r_{c}$ for the collision position and a characteristic length $r_{0}=\sqrt[6]{18C/\mu\Omega{}^{2}}$.
 Fig.~\ref{fig:collison}(a) shows  $r_{c}$ is distributed in $0\sim20$~nm for multi hard-core collsions,  
 indicationg that the max work done by RAF for one collision is on the order of tens of nK$\cdot k_{B}$.

\emph{Conclusion and discussion.---}
In summary, we proposed a novel hybrid trap consisting of a PT and modulated ODTs 
that takes advantage of the deep-potential feature
of the PT while eliminating the micromotion problem 
in the cold hybrid ion-atom system. 
The trap can suppress the micromotion temperature  to the order of nK, 
offering a new route to reach the quantum collision in the ion-atom mixture. 
Our proposed schemes are quite feasible since 
an ion trap system with high optical accessibility has been demonstrated in~\cite{he2021ion}. 
Compared with previous optical trapping methods for ions where the PT is switched off~\cite{schneider2010optical,huber2014afaroffresonance,schmidt2018optical,schmidt2020optical}, 
our scheme can stably trap elementary and molecular ions all through the cold ion-atom collision process due to the always-on PT,  
which provides an ideal platform for studying cold ion-atom collision and reaction and investigating the properties of the produced molecular ion.

\begin{acknowledgments}
This work was supported by the National Key Research and Development Program of China (No. 2017YFA0304100, 2024YFA1409403), 
the National Natural Science Foundation of China (Grant No. 11774335, No. 11734015, and No. 12204455), 
the Key Research Program of Frontier Sciences, CAS (Grant No. QYZDY-SSWSLH003), 
and the Innovation Program for Quantum Science and Technology (Grant No. 2021ZD0301604 and No. 2021ZD0301200).

J.-M. Cui and S.-J. Sun contributed equally to this work. Innovation Program for Quantum Science and Technology

The data that support the findings of this article are openly available \cite{data_availability}

\end{acknowledgments}

\appendix

\section{Ion Dynamics in the Hybrid Trap Potential}

The electrical potential of a classical Paul trap is
\begin{equation}
U_{\mathrm{Paul}}(\mathbf{r},t)=U_{E}(\mathbf{r})+\widetilde{U}_{E}(\mathbf{r})\cos(\Omega t),
\end{equation}
where $U_{E}(\mathbf{r})$ and $\widetilde{U}_{E}(\mathbf{r})$ are
the potentials generated by the DC and RF fields respectively. For
a symmetric linear Paul trap with four rods or blades, the electrical
potentials can be expressed as \citep{berkeland1998minimization}
\begin{align}
U_{E}(\mathbf{r}) & =k_{0}z^{2}-\frac{1}{2}k_{0}(x^{2}+y^{2}),\\
\widetilde{U}_{E}(\mathbf{r}) & =k_{1}(x^{2}-y^{2})+V_{RF}/2,
\end{align}
where $k_{0}=\kappa QV_{DC}/Z_{0}^{2}$, $k_{1}=QV_{RF}/2R^{2}$,
where $\kappa$ is a geometric factor of the trap, $Q$ is the charge
of the ion , $V_{DC}$ and $V_{RF}$ are the DC and RF voltages applied
on the trap, $R$ is the perpendicular distance from the trap axis
to the trap electrodes, $Z_{0}$ is the distance from the trap center
to the end-cap electrodes. The equations of motion for a single ion
of mass $m$ in the above potential are give by Mathieu equations
\begin{equation}
\ddot{u_{l}}+\left[a_{0,l}+2q_{0,l}\cos(\Omega t)\right]\frac{\Omega^{2}}{4}u_{l}=0,
\end{equation}
where $\mathbf{u}=u_{x}\hat{x}+u_{y}\hat{y}+u_{z}\hat{z}$ is the
position of the ion and
\begin{align}
a_{0,x} & =a_{0,y}=-\frac{1}{2}a_{0,z}=a_{0}=\frac{4k_{0}}{m\Omega^{2}},\\
q_{0,x} & =-q_{0,y}=q_{0}=\frac{4k_{1}}{m\Omega^{2}},\\
q_{0,z} & =0.
\end{align}

The potential generated by the hybrid trap is,
\begin{widetext}
  \begin{align}
    U(\mathbf{r},t) & =U_{E}(\mathbf{r})+\widetilde{U}_{E}(\mathbf{r})\cos(\Omega t)+\sum_{i=1}U_{D,i}(\mathbf{r})\left[1+\eta_{i}\cos(\Omega t)\right],\\
     & =\Phi_{0}(\mathbf{r})+\Phi_{1}(\mathbf{r})\cos(\Omega t)\label{eq:Potential}
  \end{align}
\end{widetext}

where $U_{D,i}(\mathbf{r})$ is the potential of $i$'th ODT, $\eta_{i}$
is the modulation depth of the ODT, $\Phi_{0}(\mathbf{r})=U_{E}(\mathbf{r})+\sum_{i}^{N}U_{D,i}(\mathbf{r})$
and $\Phi_{1}(\mathbf{r})=\widetilde{U}_{E}(\mathbf{r})+\sum_{i}^{N}\eta_{i}U_{D,i}(\mathbf{r})$
are the direct and alternating potentials respectively. If we use
the two dipole trap scheme in the main text and use the zero alternating
trapping condition, the modulation parameters will be $\eta_{1}=-\eta_{2}=1$.
The potential of ODTs are \citep{grimm2000optical}
\begin{equation}
U_{D}(\mathbf{r})=\frac{\pi c^{2}\Gamma}{2\omega_{0}^{3}}(\frac{2}{\text{\ensuremath{\Delta_{2}}}}+\frac{1}{\text{\ensuremath{\Delta_{1}}}})I(\mathbf{r}),
\end{equation}
where $I(\mathbf{r})$ is the intensity of the specified ODT, $\Gamma$
is the damping rate of $P$ state, $\Delta_{2}$ is the laser detuning
from $S_{1/2}$ to $P_{3/2}$ transition, $\Delta_{1}$ is the laser
detuning from $S_{1/2}$ to $P_{1/2}$ transition, and $\omega_{0}$
is the ODT laser frequency. The intensity profile of a focused astigmatic
Gaussian beam along axis $\hat{c}$ in coordinate $(\hat{a},\hat{b},\hat{c})$
can be written as
\begin{equation}
I(\mathbf{r})=I_{0}\frac{w_{0a}}{w_{a}(c)}\frac{w_{0b}}{w_{b}(c)}\exp\left\{ -2\left[\frac{a^{2}}{w_{a}(c)^{2}}+\frac{b^{2}}{w_{b}(c)^{2}}\right]\right\} ,
\end{equation}
where $w_{a/b}(c)=w_{0,a/b}\sqrt{1+(c/z_{Ra/b})^{2}}$ are the beam
waists for $a$ and $b$ axes, i.~e. short and long axes respectively,
$z_{Ra/b}=\pi w_{0,a/b}^{2}/\lambda$ are the Rayleigh ranges and
$\lambda$ is the ODT laser wavelength. Applying the scheme with two
orthogonal astigmatic Gaussian beam spots at $z=0$ in the main text,
the potentials are
\begin{widetext}
\begin{align}
\Phi_{0}(x,y,0) & =k_{0}z^{2}-\frac{1}{2}k_{0}(x^{2}+y^{2})+U_{0}\exp(-2x^{2}/w_{a}^{2}-2y^{2}/w_{b}^{2})+U_{0}\exp(-2x^{2}/w_{b}^{2}-2y^{2}/w_{a}^{2}),\\
\Phi_{1}(x,y,0) & =k_{1}(x^{2}-y^{2})+U_{0}\exp(-2x^{2}/w_{a}^{2}-2y^{2}/w_{b}^{2})-U_{0}\exp(-2x^{2}/w_{b}^{2}-2y^{2}/w_{a}^{2}),
\end{align}
\end{widetext}
where $U_{0}=\pi c^{2}\Gamma(\Delta_{2}+2\Delta_{1})I_{0}/2\omega_{0}^{2}\Delta_{2}\Delta_{1}$.
Expanding $\Phi_{1}(x,y,0)$ near the trap center ($x=0$, $y=0$)
and setting the alternating force $\nabla\Phi_{1}$ to zero, we can
derive the zero alternating potential condition (ZATC) 
\begin{equation}
U_{0,\mathrm{ZACT}}=\frac{k_{1}w_{a}^{2}w_{b}^{2}}{2(w_{a}^{2}-w_{b}^{2})}.\label{eq:ZACT}
\end{equation}
For the other arbitrary $U_{0}$, there will be residual alternating
forces (RAFs) near the trap center, in this case, we can define a
RAF parameter $\Delta q=4(k_{1}+k_{a}-k_{b})/m\Omega^{2}$ (where
$k_{a,b}=2U_{0}/w_{a,b}^{2}$, see the subsection \ref{subsec:Approx_eMM}).
Comparing the original Paul trap, a ratio of RAF of the hybrid trap
can be defined as $\beta=\Delta q/q_{0}$.

To study three-dimensional dynamics of a single ion in the hybrid
trap, we can perform numerical simulation by reserving the full form
of the potentials. We take the equation along the $x$ axis for example.
From the equation of motion driven by forces 
\begin{equation}
m\ddot{x}=-\frac{\partial U(\mathbf{r},t)}{\partial x}=-\frac{\partial\Phi_{0}}{\partial x}-\frac{\partial\Phi_{1}}{\partial x}\cos(\Omega t),
\end{equation}
we can get
\begin{equation}
\ddot{x}+\left[a_{x}(x,y,z)+2q_{x}(x,y,z)\cos\left(\Omega t\right)\right]\frac{\Omega^{2}}{4}x=0,\label{eq:PDME}
\end{equation}
where 
\begin{align*}
a_{x}(x,y,z) & =\frac{4}{m\Omega^{2}}\frac{\partial\Phi_{0}}{x\partial x},\\
q_{x}(x,y,z) & =\frac{2}{m\Omega^{2}}\frac{\partial\Phi_{1}}{x\partial x},
\end{align*}
are Mathieu equation parameters that depend on the ion's position.
The position dependent Mathieu equation parameters in three dimensions
can be deduced as
\begin{widetext}
\begin{align}
a_{x}(x,y,z)= & \frac{4}{m\Omega^{2}}\left[-k_{0}-U_{0}g(z)\left(f_{1}(x,y,z)\frac{4}{w_{a}^{2}(z)}+f_{2}(x,y,z)\frac{4}{w_{b}^{2}(z)}\right)\right],\nonumber \\
a_{y}(x,y,z)= & \frac{4}{m\Omega^{2}}\left[-k_{0}-U_{0}g(z)\left(f_{1}(x,y,z)\frac{4}{w_{b}^{2}(z)}+f_{2}(x,y,z)\frac{4}{w_{a}^{2}(z)}\right)\right],\label{eq:al}\\
a_{z}(x,y,z)= & \frac{4}{m\Omega^{2}}\left[2k_{0}+U_{0}g(z)\left(f_{1}M_{1}+f_{2}M_{2}\right)+U_{0}N(z)\left(f_{1}+f_{2}\right)\right],\nonumber 
\end{align}
and
\begin{align}
q_{x}(x,y,z) & =\frac{2}{m\Omega^{2}}\left[2k_{1}+U_{0}g(z)\left(f_{1}(x,y,z)\frac{4}{w_{a}^{2}(z)}-f_{2}(x,y,z)\frac{4}{w_{b}^{2}(z)}\right)\right],\nonumber \\
q_{y}(x,y,z) & =\frac{2}{m\Omega^{2}}\left[-2k_{1}+U_{0}g(z)\left(f_{1}(x,y,z)\frac{4}{w_{b}^{2}(z)}-f_{2}(x,y,z)\frac{4}{w_{a}^{2}(z)}\right)\right],\label{eq:ql}\\
q_{z}(x,y,z) & =\frac{2}{m\Omega^{2}}\left[U_{0}g(z)\left(f_{1}M_{1}-f_{2}M_{2}\right)+U_{0}N(z)\left(f_{1}-f_{2}\right)\right],\nonumber 
\end{align}
\end{widetext}
where 
\begin{align}
f_{1}(x,y,z)= & \exp\left[-2x^{2}/w_{a}^{2}(z)-2y^{2}/w_{b}^{2}(z)\right]\\
f_{2}(x,y,z)= & \exp\left[-2x^{2}/w_{b}^{2}(z)-2y^{2}/w_{a}^{2}(z)\right]\\
g(z)= & \frac{1}{\sqrt{1+(z/z_{Ra})^{2}}\sqrt{1+(z/z_{Rb})^{2}}}
\end{align}
\begin{align}
N(z)= & -g(z)\left(\frac{1}{z_{Ra}^{2}+z^{2}}+\frac{1}{z_{Rb}^{2}+z^{2}}\right)\\
M_{1}(x,y,z)= & \frac{4x^{2}}{w_{0,a}^{2}z_{Ra}^{2}\alpha_{a}^{4}(z)}+\frac{4y^{2}}{w_{0,b}^{2}z_{Rb}^{2}\alpha_{b}^{4}(z)}\\
M_{2}(x,y,z)= & \frac{4x^{2}}{w_{0,b}^{2}z_{Rb}^{2}\alpha_{b}^{4}(z)}+\frac{4y^{2}}{w_{0,a}^{2}z_{Ra}^{2}\alpha_{a}^{4}(z)}
\end{align}
and $w_{a/b}^{2}(z)=w_{0,a/b}^{2}\alpha_{a/b}^{2}(z)$,$\alpha_{a/b}(z)=\sqrt{1+(z/z_{Ra/b})^{2}}$,
$z_{Ra/b}=\pi w_{0,a/b}^{2}/\lambda$.

In the presence of a stray electrical field $\mathbf{E}=E_{x}\hat{x}+E_{y}\hat{y}+E_{z}\hat{z}$,
the equations of motion will be 
\begin{align}
\ddot{x}+\left[a_{x}(x,y,z)+2q_{x}(x,y,z)\cos\left(\Omega t\right)\right]\frac{\Omega^{2}}{4}x & =\frac{Q\cdot E_{x}}{m},\\
\ddot{y}+\left[a_{y}(x,y,z)+2q_{y}(x,y,z)\cos\left(\Omega t\right)\right]\frac{\Omega^{2}}{4}y & =\frac{Q\cdot E_{y}}{m},\\
\ddot{z}+\left[a_{z}(x,y,z)+2q_{z}(x,y,z)\cos\left(\Omega t\right)\right]\frac{\Omega^{2}}{4}z & =\frac{Q\cdot E_{z}}{m}.
\end{align}

\begin{table}
\caption[Parameters of Paul trap and hybrid trap in this work]{\label{tab:S1} Numerical simulation parameters for trapping $^{171}\mathrm{Yb}^{+}$.
The shared trap parameters with $\Omega=2\pi\times2$~MHz (RF frequency),
$\kappa=0.6$, $V_{DC}=1$~V, $V_{RF}=10$~V, $R=0.5$~mm and $Z_{0}=1$~mm. }

\begin{ruledtabular}
\begin{tabular}{ccc}
 & Paul Trap & Hybrid Trap\tabularnewline
\hline 
$a_{x}$ & -8.575$\times10^{-3}$ & \tabularnewline
$a_{y}$ & -8.575$\times10^{-3}$ & \tabularnewline
$a_{z}$ & 17.15$\times10^{-3}$ & \tabularnewline
$q_{x}$ & 0.286 & \tabularnewline
$q_{y}$ & -0.286 & \tabularnewline
$q_{z}$ & 0 & \tabularnewline
$\omega_{x}/2\pi$ & 179.7 kHz & 840.2 kHz\tabularnewline
$\omega_{y}/2\pi$ & 179.7 kHz & 840.2 kHz\tabularnewline
$\omega_{z}/2\pi$ & 131.0 kHz & 131.0 kHz\tabularnewline
\end{tabular}
\end{ruledtabular}

\end{table}

\section{Ion Micromotion temperature}

\subsection{Micromotion temperature of a trapped ion}

The Mathieu equation when DC field $\mathbf{E}$ shifts the equilibrium
position of the ion writes \citet{berkeland1998minimization} 
\[
\ddot{u_{l}}+\left[a_{l}+2q_{l}\cos(\Omega t)\right]\frac{\Omega^{2}}{4}u_{l}=\frac{Q\cdot E_{l}}{m},
\]
and the solution is
\[
u_{l}(t)=\left(u_{0,l}+u_{1,l}\cos(\omega_{l}t+\varphi_{l})\right)\left(1+\frac{q_{l}}{2}\cos(\Omega t)\right),
\]
where the equilibrium position 
\begin{equation}
u_{0,l}\cong\frac{4Q\cdot E_{l}}{m\Omega^{2}(a_{l}+q_{l}^{2}/2)},
\end{equation}
and $u_{1,l}$ is the secular motion amplitude. The average kinetic
energy due to motion along $\hat{u}_{l}$ is
\begin{align}
E_{k,l} & =\frac{1}{4}mu_{1,l}^{2}(\omega_{l}^{2}+\frac{1}{8}q_{l}^{2}\Omega^{2})+\frac{m\Omega^{2}}{16}q_{l}^{2}u_{0,l}^{2},\nonumber \\
 & =\frac{1}{4}mu_{1,l}^{2}\omega_{l}^{2}+\frac{1}{32}mq_{l}^{2}u_{1,l}^{2}\Omega^{2}+\frac{m\Omega^{2}}{16}q_{l}^{2}u_{0,l}^{2}\nonumber \\
 & =E_{0,l}+E_{\mathrm{iMM},l}+E_{\mathrm{eMM},l}\label{eq:Kinetic}
\end{align}
where the second and third term represent the kinetic energy due to
intrinsic micromotion (iMM) and excess micromotion (eMM) along $\hat{u}_{l}$,
respectively. So the temperature caused by the eMM in all directions
reads
\begin{equation}
T_{\mathrm{eMM}}=\sum_{l}\frac{E_{\mathrm{eMM},l}}{k_{B}}=\frac{m\Omega^{2}}{16k_{B}}\sum_{l}q_{l}^{2}u_{0,l}^{2}.\label{eq:T_M98}
\end{equation}
$T_{\mathrm{eMM}}$ represents the lowest attainable temperature caused
by excess micromotion of an ion in the trap, it can be zero in theory
by setting $u_{0,l}=0$, however experimental technics on adjusting
the voltages on DC electrodes limit achievable $T_{\mathrm{eMM}}$.
The temperature caused by the iMM is 
\begin{equation}
T_{\mathrm{iMM}}=\sum_{l}\frac{E_{\mathrm{iMM},l}}{k_{B}}=\frac{m\Omega^{2}}{32k_{B}}\sum_{l}q_{l}^{2}u_{1,l}^{2}.\label{eq:T_iMM}
\end{equation}
Even in the condition that the secular motion is cooled to its ground
state (when $u_{1,l}=\sqrt{\hbar/2m\omega_{l}}$) and the eMM is well
compensated, the ion still has a temperature limit caused by the iMM
\begin{equation}
T_{0,\mathrm{iMM}}=\frac{\hbar\Omega^{2}}{64k_{B}}\sum_{l}\frac{q_{l}^{2}}{\omega_{l}}.\label{eq:T_iMM0}
\end{equation}
With trap parameters in Tab.~\ref{tab:S1}, the $T_{0,\mathrm{iMM}}$
of the Paul trap is 2.73~$\mu$K. Considering a ratio of RAF $\beta=\Delta q/q_{0}=1\%$
for a hybrid trap, which is achievable in experiment, the corresponding
$T_{0,\mathrm{iMM}}$ is 58.3~pK. 

Now we consider the case that the Mathieu equation parameters are
in a position-dependent form: $a_{l}(\mathbf{u})$, $q_{l}(\mathbf{u})$.
If the ion oscillates near $\mathbf{u}_{0}$ in the hybrid trap under
a stray electrical field $\mathbf{E}$, and it has been cooled to
a low temperature $\left|\mathbf{u}-\mathbf{u}_{0}\right|\ll w_{a}$,
then the Mathieu equation can be approximated as 
\begin{equation}
\ddot{u_{l}}+\left[a_{l}(\mathbf{u}_{0})+2q_{l}(\mathbf{u}_{0})\cos\left(\Omega t\right)\right]\frac{\Omega^{2}}{4}u_{l}=\frac{Q\cdot E_{l}}{m},\label{eq:DC_Motion}
\end{equation}
where the equilibrium position $\mathbf{u}_{0}$ can be numerically
solved by iterating
\[
u_{0,l}^{i+1}=\frac{4Q\cdot E_{l}}{m\Omega^{2}\left[a_{l}(\mathbf{u}_{0}^{i})+q_{l}^{2}(\mathbf{u}_{0}^{i})/2\right]},
\]
with an initial position $\mathbf{u}_{0}^{0}=(0,0,0)$. Once the equilibrium
position $\mathbf{u}_{0}$ is solved, we can get the temperature caused
by excess micromotion
\begin{equation}
T_{\mathrm{eMM}}=\frac{m\Omega^{2}}{16k_{B}}\sum_{l}q_{l}^{2}(\mathbf{u}_{0})u_{0,l}^{2}.\label{eq:T_M_main}
\end{equation}

\subsection{Approximated ion's eMM temperature expression \label{subsec:Approx_eMM}}

Take Paul trap center as the origin of coordinate, supposing we use
full modulation, and the misalignment between the Paul trap and the
ODT is $\delta x$. We define $\Delta x=x-\delta x$, then we can
get the approximation at $(x,0,0)$
\begin{align}
a_{x}(x) & =\frac{4}{m\Omega^{2}}\left[-k_{0}-U_{0}\left(e^{-2\Delta x^{2}/w_{a}^{2}}\frac{4}{w_{a}^{2}}+e^{-2\Delta x^{2}/w_{b}^{2}}\frac{4}{w_{b}^{2}}\right)\right] \nonumber\\
 & \approx-\frac{4}{m\Omega^{2}}\left[k_{0}+2k_{a}+2k_{b}-4\left(k_{a}/w_{a}^{2}+k_{b}/w_{b}^{2}\right)\Delta x^{2}\right] \nonumber\\
 & =a-(x-\delta x)^{2}/x_{a}^{2}, \label{eq:a_approx}
\end{align}
\begin{align}
q_{x}(x) & =\frac{2}{m\Omega^{2}}\left[2k_{1}+U_{0}\left(e^{-2\Delta x^{2}/w_{a}^{2}}\frac{4}{w_{a}^{2}}-e^{-2\Delta x^{2}/w_{b}^{2}}\frac{4}{w_{b}^{2}}\right)\right]\nonumber \\
 & \approx\frac{4}{m\Omega^{2}}\left[k_{1}+\left(k_{a}-k_{b}\right)+2\left(k_{b}/w_{b}^{2}-k_{a}/w_{a}^{2}\right)\Delta x^{2}\right]\nonumber \\
 & =\Delta q+(x-\delta x){}^{2}/x_{b}^{2},\label{eq:q_approx}
\end{align}
where $k_{a,b}=2U_{0}/w_{a,b}^{2}$, $a=-4(k_{0}+2k_{a}+2k_{b})/m\Omega^{2}$, $\Delta q=4(k_{1}+k_{a}-k_{b})/m\Omega^{2}$,
$x_{a}^{2}=-m\Omega^{2}/16(k_{a}/w_{a}^{2}+k_{b}/w_{b}^{2})$ and
$x_{b}^{2}=m\Omega^{2}/8(k_{b}/w_{b}^{2}-k_{a}/w_{a}^{2})$. The
trap center will shift to an equilibrium position under a stray field
$E$ 
\begin{align*}
x_{0} & =\frac{4Q\cdot E}{m\Omega^{2}\left(a+\Delta q^{2}/2\right)}\approx\frac{4Q\cdot E}{m\Omega^{2}a}=-\frac{Q\cdot E}{k_{0}+2k_{a}+2k_{b}}.
\end{align*}
 Combined with Eq.~\ref{eq:T_M98}, the approximated eMM temperature
expression in this situation reads
\begin{align*}
T_{\mathrm{eMM}} & \approx\frac{m\Omega^{2}}{16k_{B}}q_{x}^{2}(x_{0},0,0)x_{0}^{2}\\
 & \approx\frac{m\Omega^{2}}{16k_{B}}\left(\Delta q+\frac{(x_{0}-\delta x){}^{2}}{x_{b}^{2}}\right)^{2}x_{0}^{2},
\end{align*}
which clearly illustrates the roles that imperfections $\Delta q$,
$\delta x$ play in restricting the lowest attainable temperature.
In this expression, we approximate the iterated equilibrium position
$u_{0}$ to $x_{0},$which represents the equilibrium position when
no imperfections are presented. If we want the expression to be more
precise, we could simply substitute $x_{0}$ into the right side of
the iteration formula in Eq.~\ref{eq:T_M}, and then  let $u_{0}$
equal to the first iteration solution $x_{1}=4Q\cdot E/\left[m\Omega^{2}\left(a_{x}(x_{0},0,0)+q_{x}^{2}(x_{0},0,0)/2\right)\right]$,
which gives
\[
T_{\mathrm{eMM}}\approx\frac{m\Omega^{2}}{16k_{B}}\left(\Delta q+\frac{(x_{1}-\delta x){}^{2}}{x_{b}^{2}}\right)^{2}x_{1}^{2}.
\]

\section{Numerical Simulation of Ion-Atom Collisions in three Dimensions}

In this section a numerical simulation method used for calculating
the classical low-energy three-dimensional ion-atom collision trajectory
is introduced. We simulated and compared two situations: the collision
happens in a normal Paul trap and the hybrid trap we proposed, respectively.
In the following content, we introduce how to calculate the collision
process that happened in the hybrid trap, and the situation that happened
in the Paul trap can be similarly and easily deduced. 

% \begin{figure}
% \centering\includegraphics[width=9cm]{FigSM1}
% \caption[Numerical 3D simulations of ion-atom collisions]{
% \label{fig:S:Approching3D_xyz} 
% Numerical 3D simulations of ion-atom
% ($^{87}\mathrm{Rb}$ and $^{172}\mathrm{Yb}^{+}$) collisions in a
% hybrid trap (a-c) and a Paul trap (d-f), respectively.}
% \end{figure}
An $^{87}\mathrm{Rb}$ atom of mass $m_{a}$ approaches an $^{172}\mathrm{Yb^{+}}$
ion of mass $m_{i}$ (initially resting in the center of the hybrid
trap) from an arbitrary position (in the following content, we take
the initial position as $x=100$ nm, $y=100$ nm, $z=100$ nm without
loss of generality), by the attractive force between the two, the
potential of which originates from the interaction between the ion
and polarized atom and thus has the form $-C_{4}/r^{4}$, where $C_{4}$
represent the polarizability of atom and the value of $C_{4}$ is
taken from Ref. \citep{tomza2019coldhybrid}. For our selected $^{87}\mathrm{Rb}$
-$^{172}\mathrm{Yb^{+}}$ pair, $C_{4}=160E_{h}a_{0}^{4}$, where
$E_{h}$ is the Hartree energy and $a_{0}$is the Bohr radius. At
short ranges, we add a repulsive $r^{-6}$ term to simulate a hard-core
potential, where the hard-core collision distance is on the order
of nm when the coefficient $C_{6}$ is given as a fraction of $C_{4}$
as $C_{6}=(10a_{0})^{2}\times C_{4}$ in our simulation. Therefore,
the total ion-atom potential reads
\[
V(r)=-C_{4}/r^{4}+C_{6}/r^{6}.
\]

Since we intentionally select the magic wavelength $\lambda=$422.3\ nm
of the $^{87}\mathrm{Rb}$ atom as the ODT laser wavelength, the only
potential experienced by the atom when approaching or moving away
is the $V(R)$ above. Thus, the equation of motion of the atom reads
\begin{equation}
m_{a}\frac{d^{2}\mathbf{r_{a}}}{dt^{2}}=-\frac{4C_{4}}{(\mathbf{r_{a}}-\mathbf{r_{i}})^{6}}(\mathbf{r_{a}}-\mathbf{r_{i}})+\frac{6C_{6}}{(\mathbf{r_{a}}-\mathbf{r_{i}})^{8}}(\mathbf{r_{a}}-\mathbf{r_{i}}),\label{eq:eom_atom-1}
\end{equation}
where $\mathbf{r_{a}}=(x_{a},y_{a},z_{a})$ and $\mathbf{r_{i}}=(x_{i},y_{i},z_{i})$
are the 3D positions of atom and ion separately. 

Combining Eq.~\ref{eq:PDME} and the attractive force from the polarized
atom, the equation of motion of the ion along the $x$ direction is
\begin{widetext}
  \begin{equation}
    m_{i}\frac{d^{2}x_{i}}{dt^{2}}=-\frac{m_{i}\Omega^{2}}{4}\left[a_{x}(r_{i})+2q_{x}(r_{i})\cos(\Omega t+\varphi)\right]x_{i}+\frac{4C_{4}}{(r_{a}-r_{i})^{6}}(x_{a}-x_{i}),\label{eq:eom_ion-1}
  \end{equation}
\end{widetext}

and the equations of motion of the ion along the $y$ and $z$ direction
can be written down similarly.

We use the explicit Runge-Kutta method of order eight to solve the
equations of motion by combining Eq.~\ref{eq:al}, Eq.~\ref{eq:ql},
Eq.~\ref{eq:eom_atom-1} and Eq.~\ref{eq:eom_ion-1}. Fig.~\ref{fig:collison}
shows a simulated collision trajectory of an atom-ion pair in $x$,
$y$ and $z$ directions respectively. The lifetime of the bound atom-ion
pair produced in the hybrid trap is significantly prolonged due to
the low RF heating near the hybrid trap center.

\section{Work Done by RF Field on the Ion}

When an ion moves in a Paul trap without any disturbance, the RF field
does zero work on the ion for an RF period. However, in a cold hybrid
ion-atom system, the polarized atom will exert an attraction force
on the ion (by attraction potential $V(r)=-C_{4}/r^{4}$) and thus
disturb the conserved cycle of energy transfer, and finally, the ion-atom
system gains net energy from the RF field, which increases the temperature
of the system. Here, we calculate the work done by the RF field on
the ion when an atom approaches the ion and the collision happens.
Suppose the collision point is at $r_{c}$, and the collision happens
at $t=0$. When the attraction force between the ion-atom pair dominates,
the distance between the ion and atom follows $r(t)=(18C_{4}/\mu)^{1/6}|t|^{1/3},$so
the trajectory of the ion becomes $r_{i}(t)=r_{c}-\mu r(t)/m_{i},$
where $\mu=m_{i}m_{a}/(m_{i}+m_{a})$ is the reduced mass of the ion-atom
pair. We can write down the work done by the RF field on the ion when
the atom is close enough to perturb the motion of ion in the trap
\[
W=\int_{-t_{1}}^{t_{1}}2q(r_{i})\cdot\frac{m_{i}\Omega^{2}}{4}r_{i}\cos(\Omega t+\phi)\dot{r_{i}}\mathrm{d}t,
\]
where integration range $t=-t_{1}\sim t_{1}$ is the time when the
perturbation from the atom dominates, and we can take $t_{1}\approx0.8$
for approximation\citep{cetina2012PRL}. By substituting
Eq.~\ref{eq:q_approx} and the trajectory of the ion, we get the
work done by the RF field during one collision
\[
W=W_{1}+W_{2},
\]
\[
W_{1}=\frac{1}{3}\mu\Omega^{2}R^{2}\sin\phi\cdot\Delta q\int_{0}^{\Omega t_{1}}(A-B\tau^{1/3})\cdot\tau^{-2/3}\sin\tau\cdot d\tau,
\]
\[
W_{2}=\frac{1}{3}\mu\Omega^{2}R^{2}\sin\phi\cdot\frac{R^{4}}{r_{0}^{2}x_{b}^{2}}\int_{0}^{\Omega t_{1}}(A-B\tau^{1/3})^{3}\cdot\tau^{-2/3}\sin\tau\cdot d\tau,
\]
where $A=r_{c}r_{0}/R^{2}$, $B=\mu r_{0}^{2}/m_{i}R^{2}$, $r_{0}=\sqrt[6]{18C_{4}/\mu\Omega{}^{2}}$.
$W_{1}$ comes from the uncompensated alternating force $\Delta q$
and $W_{2}$ comes from the dependence of $q$ on position. For the
first collision, when the atom approaches the ion from infinity, the
collision happens at $r_{c,0}=1.11(m_{a}/m_{i})^{5/6}R\approx20.8$~nm
\citep{cetina2012PRL}, and the numerical results give
the max work heating $W_{1}/k_{B}\approx-85.2$~nK, $W_{2}/k_{B}\approx27.8$~nK
for $\Delta q=-0.001$ at $\sin\phi=1$. Here we intentionally choose
$\Delta q$ as negative because we want $q(r)$ to be nearly $0$
at the collision points, not at the origin. 

\begin{figure}
\centering\includegraphics[width=8.5cm]{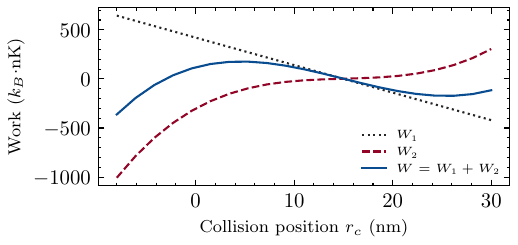}
\caption[Maximum work done by the RF field during one collision]{
    \label{fig:work}Maximum work done by the RF field during one collision
at $\sin\phi=1.$}
\end{figure}
After the first collision happens at $r_{c,0}$, the following collision
position $r_{c,i}$ for the $i$'th collision is unpredictable. From
Fig.~\ref{fig:collison}(a,b) we see that the collisions
most likely happen at approximately $0\sim20$~nm at micromotion
directions ($x$ and $y$ axis). The Fig.~\ref{fig:work} illustrates
the relation of $W=W_{1}+W_{2}$ with $r_{c}$ for $\Delta q=-0.001$,
from which we could see that by compensating the $\Delta q$ nearly
zero, the work done by residual RF field during one collision is on
the order of tens of nK$\cdot k_{B}$. Therefore, the relative energy
between an ion-atom pair could easily run below the characteristic
energy scale $E_{k}$ after one or more collisions, making the formation
of a bound state (like a molecular ion) possible.

\bibliography{HybridTrap}

\end{document}